# ADVANCES IN THE TREATMENT OF TRIMMED CAD MODELS DUE TO ISOGEOMETRIC ANALYSIS

BENJAMIN MARUSSIG[*,†]

[*]Graz University of Technology
Graz Center of Computational Engineering (GCCE)
Krenngasse 37/I, 8010 Graz, Austria
marussig@tugraz.at; www.gcce.tugraz.at

[†]Graz University of Technology
Institute of Mechanics
Kopernikusgasse 24/IV, 8010 Graz, Austria

**Abstract.** Trimming is a core technique in geometric modeling. Unfortunately, the resulting objects do not take the requirements of numerical simulations into account and yield various problems. This paper outlines principal issues of trimmed models and highlights different analysis-suitable strategies to address them. It is discussed that these concepts not only provide important computational tools for isogeometric analysis, but can also improve the treatment of trimmed models in a design context.

## 1 INTRODUCTION

Isogeometric analysis (IGA) aims to bridge the gap between computer aided geometric design (CAGD) and analysis by using CAGD technologies for numerical simulations. Since the introduction of IGA in 2005, it has been demonstrated that the synthesis of these disciplines allows not only an improved interaction, but yields many computational advantages[1–3]. Nowadays, IGA is widely recognized as a powerful alternative to the conventional analysis methodology. In the following, the attention is drawn to a somewhat different aspect, namely (potential) benefits for CAGD due to developments made in IGA. It is focused on challenges concerning robustness and interoperability. In particular, the treatment of trimmed models is addressed, because these representations play a central role in engineering design and the integration of design and analysis[4]. First, the evolution of trimmed CAD models is presented in order to outline the related problems. Based on that, corresponding advances of IGA are discussed.

## 2   A BRIEF HISTORY OF TRIMMED SOLID MODELS

The problem of computing surface-to-surface intersections (SSI) is closely related to trimming and thus, it is discussed at the beginning of this section. Then, the formulation of solid models defined by trimmed surfaces is presented and finally related robustness issues and the role of trimmed models with respect to the exchange of CAD data are discussed.

### 2.1   SSI operations

Computing intersections of surfaces is a crucial task in various types of modeling processes. First of all, it is the core ingredient for Boolean operations which are the most important functions in creating CAD objects[5]. In general, the intersection of two parametric surfaces

$$\boldsymbol{S}_1(u,v) = (x_1(u,v), y_1(u,v), z_1(u,v)) \quad (1)$$

$$\boldsymbol{S}_2(s,t) = (x_2(s,t), y_2(s,t), z_2(s,t)) \quad (2)$$

leads to a system of three nonlinear equations[6]. These equations represent the three coordinate differences of the surfaces, $\boldsymbol{S}_1$ and $\boldsymbol{S}_2$, with the four unknown surface parameters $u, v, s, t$. In most cases, the solution describes a curve, but intersection points, subsurfaces, or empty sets may occur as well.

Efficiently providing *all* features of these solutions is the purpose of SSI operations[7]. The development of a good SSI procedure is a very challenging task due to the fact that the operation has to be *accurate*, *efficient*, and *robust*. These attributes are indeed quite contradictory and the definition of an adequate balance between them depends strongly on the application context. Early solid modeling systems employed analytic methods to compute exact parametric descriptions of intersections between linear and quadratic surfaces[8]. Unfortunately, the algebraic complexity of an intersection increases rapidly with the degree of $\boldsymbol{S}_1$ and $\boldsymbol{S}_2$, which has been thoroughly discussed by Sederberg and co-workers[9–11] in the 1980s. This makes analytic approaches impractical; a fact often illustrated by the algebraic degree of an intersection of two general bicubic surfaces which is 324.

Hence, alternative SSI schemes are needed. These concepts can be broadly classified as lattice evaluation schemes[12,13], subdivision methods[14], and marching methods[15,16]. The former reduces the dimensionality of the problem by computing intersections of a number of isocurves of $\boldsymbol{S}_1$ with $\boldsymbol{S}_2$ and vice versa. The second strategy uses approximations of the actual surfaces, often defined by a set of piecewise linear elements, and computes the related intersections with respect to the simplified objects. Finally, marching methods define an intersection curve by stepping piecewise along the curve. This requires detection of appropriate starting points, determination of point sequences along the intersection that emit from the starting points, and proper sorting and merging of these individual sequences. Marching methods are by far the most widely used schemes due to their generality and ease of implementation[17]. However, each intersection strategy has its advantages and drawbacks, hence SSI algorithms usually use hybrid concepts that combine different features of these approaches[17–19].

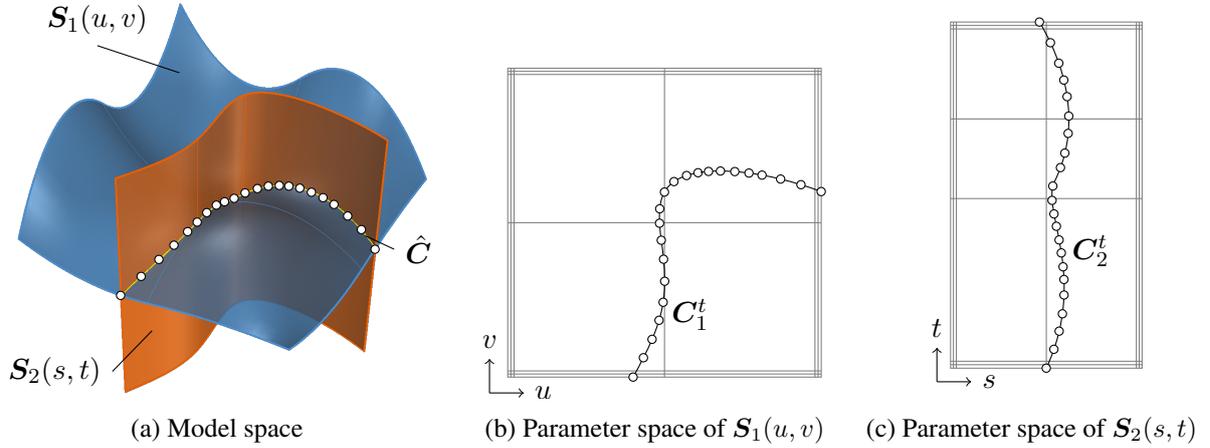

(a) Model space  (b) Parameter space of $S_1(u,v)$  (c) Parameter space of $S_2(s,t)$

Figure 1: Independent curve interpolation of an ordered point set to obtain approximations of the intersection of two patches $S_1(u,v)$ and $S_2(s,t)$. The set of sampling points depends on the SSI algorithm applied. The subsequent interpolation of these points is performed in (a) the model space and the parameter space of (b) $S_1(u,v)$ and (c) $S_2(s,t)$ leading to independent curves $\hat{C}$, $C_1^t$, and $C_2^t$.

Irrespective of the scheme applied, the initial result of a SSI operation is usually a set of sampling points that represent the intersection[20]. An *approximate* intersection curve in model space $\hat{C}$ is subsequently obtained by some curve-fitting technique such as point interpolation or least-squares approximation. Thus, $\hat{C}$ does not lie on either of the intersecting surfaces in general. Furthermore, the sampling points are mapped into the parameter spaces of $S_1$ and $S_2$, where they are again used as input for a curve-fitting procedure. This yields the main result of the SSI process, namely *trimming curves* $C^t$ in the two-dimensional parametric domains. These $C^t$ are usually represented by spline curves. They are essential because they allow the definition of arbitrarily shaped partitions within a tensor product surface, which enables proper visualization of intersecting surfaces and the application of Boolean operations. Every $C^t$ can be mapped into model space, but the resulting image $\widetilde{C}^t$ will not coincide with $\hat{C}$. In short, SSI operations yield various independent approximations of the actual intersection (see Figure 1), rather than an unambiguous solution. It is emphasized that there is no direct mapping between these different approximations and that the sampling point data for their construction is usually discarded once the curves are computed.

## 2.2  Trimmed solid models

There are various approaches for representing geometric objects[20–22]. The most popular one in engineering design is the boundary representation (B-Rep) and the benefits of storing an object's shape by means of its boundary were already elaborated in the seminal work of Braid[23]. B-Rep solid modeling utilizes SSI schemes to create arbitrarily defined free-form geometric entities. The corresponding algorithms, however, require more than the computation of intersection curves. Essential attributes of geometric modeling operators are[24]:

- the determination of the geometric surface descriptions,
- the determination of the topological descriptions, and
- the guarantee that the geometry corresponds unambiguously to the topology.

Topological data is not metrical, but addresses connectivity and dimensional continuity of a model[20]. Its determination requires the classification of the neighborhood of various entities (faces, edges, and vertices) involved in the intersections[21]. In CAGD, the term *solid* model emphasizes that a representation contains the descriptions of an object's shape, i.e., the geometry, as well as its structure, i.e., the topology; it does not refer to the dimension of the object defined.

The idea of a trimmed model appeared already in 1974 and was proposed by Pierre Bézier[25]. However, the approach was presented with little theoretical support and it took some time to develop a rigorous way to represent trimmed free-form solid models. The first formulation supporting Boolean operations and free-form geometry was presented by Farouki[26] as well as Casale and Bobrow[27,28] in the late 1980s. In general, the connectivity between intersecting surfaces is established by assigning the approximate intersection curves (which do not coincide) to a single topological entity. Further, Boolean operations define the relation of the faces, edges, and vertices of a model. Various data structures for B-Reps have been proposed to find a compromise between storage requirements and response to topological questions. The crucial discrepancy, which still exits, is that solid modeling is concerned with the use of *unambiguous* representations, but SSI schemes introduce approximations and do not provide a unique representation of an intersection. In other words, all these modeling approaches have to deal with imprecise data and thus, fail to guaranty exact topological consistency[29]. Thus, the robustness of a trimmed B-Rep becomes a crucial factor.

## 2.3 Robustness issues of trimmed models

Several robustness issues arise in case of imprecise geometric operations. As a matter of fact, numerical output from simple geometric operations can already be quite inaccurate - even for linear elements[21]. For SSI schemes, ill-conditioned intersection problems are particularly troublesome. Such cases occur when intersections are tangential or surfaces overlap, for instance. Since geometrical decisions are based on *approximate* data and arithmetic operations of *limited* precision, there is an interval of uncertainty in which the numerical data cannot yield further information[30] and the fact that SSI operations do not provide a unique intersection curve makes the situation even more delicate.

The most common strategy to address robustness issues is the use of tolerances[14]. They shall assess the quality of geometrical operations and may be adaptively defined[31,32] or dynamically updated[33]. Alternative approaches employ interval arithmetic[34] or exact arithmetic[35], but these concepts have certain drawbacks (especially with respect to efficiency) and hence, tolerance based approaches are usually preferred. Unfortunately, tolerances cannot guarantee robust algorithms since they do not deal with the inherent problem of limited-precision arithmetic.

Overall, the formulation of *robust* solid models with trimmed patches is still an open issue. This is particularly true when a model shall be transferred from a CAD system to another software tool. Since there is no canonical representation of trimmed solid models, different systems may employ different data structures and robustness checks. Consequently, data exchange involves a translation process which can lead to misinterpretation. This makes the treatment of trimmed solid models a key aspect for the interoperability of design and analysis.

## 3 DEVELOPMENTS IN ISOGEOMETRIC ANALYSIS

Since the introduction of IGA, more and more scientists in the field of computational mechanics have become aware of the advantages and deficiencies of design models and various analysis-suitable approaches dealing with CAD-related challenges have been proposed. Here, we highlight advances made in the context of local refinement of multivariate splines, which are important to derive watertight models, and the treatment of trimmed geometries.

### 3.1 Local refinement

The lack of local refinement of conventional tensor product splines was one of the first issues tackled by the IGA community. The topic emerged to an active area of research and several techniques have been developed, such as T-splines[36,37], LR-B-splines[38,39], hierarchical B-splines[40–42], and truncated hierarchical B-splines[43]. Some of these concepts were first presented in the context of CAGD (e.g., T-splines and hierarchical B-splines). However, their application in an analysis setting has provided a huge impetus to their further enhancement. In fact, these concepts have become so technically mature that the question is no longer if local refinement of multivariate spline is feasible, but what technique do you prefer.

Besides the apparent computational benefits, these advances in local refinement techniques also offer new possibilities for the design community. Admittedly, these local refinement concepts are usually not incorporated in current CAD systems (yet), but a strong indicator for the impact of IGA is a novel capability of the next version of the Standard for the Exchange of Product Model Data (STEP) – the most involved neutral exchange standard. That is, it will include entities that facilitate a canonical representation of locally refined tensor product splines[44]. To be precise, this feature affects the part "geometric and topological representations," which focuses on the definition of geometric models and represents a core component of STEP. Regarding trimmed models, the ability of local refinement can be a powerful tool as well. For instance, effects of trimming may be localized[45] or trimmed surfaces may even be joined as it is done during the conversion of trimmed B-Reps to watertight T-spline models[46].

## 3.2 Dealing with non-watertight representations

The term "non-watertight" is commonly used to stress that trimmed models have small gaps and overlaps between their intersecting surfaces. They occur due to the inevitable approximations introduced by SSI operations as discussed in Section 2.1. Watertight representations, on the other hand, possess unambiguously-defined edges and a direct link between adjacent elements. This link is missing in case of trimmed models and has to be established (or at least taken into account) in order to make them analysis-suitable. Current attempts for the integration of trimmed geometries into IGA can be divided into global and local approaches[4]. The former aims to convert trimmed solid models to watertight ones in a pre-processing step (or even already during the design stage), whereas the latter intends to enhance the simulation tool so that it is able to cope with the models' flaws.

### 3.2.1 Local approaches

The basic idea of local approaches is that trimmed parameter spaces are used as background parameterization for the simulation. Hence, there is a close relation to fictitious domain methods and the corresponding challenges are indeed similar: First, the elements needed for the analysis have to be detected[47–49]. Second, special integration techniques for elements cut by a trimming curve[47–54] have to be employed. Third, weak enforcement of boundary conditions or weak coupling of adjacent surfaces has to be addressed[50,53,55,56]. Finally, stability issues of cut elements with small support should be taken into account[45,57]. The main difference to fictitious domain methods is that an additional effort is required to associate the degrees of freedom of adjacent patches, keeping in mind that their intersections have non-matching parameterizations, gaps, and overlaps. Usually, point inversion algorithms[58,59] are utilized to establish a link between adjacent surfaces. Alternatively, simulation methods that allow discontinuities between elements[57,60] can be applied.

The majority of the publications on IGA with trimmed geometries employs such local concepts. A possible reason could be that these approaches focus on analysis aspects and thus, may seem more feasible for researchers in the field of computational mechanics. On the other hand, this also means that the number of subjects that may affect CAGD is relatively small. The essential common ground is the problem of finding robust procedures and the use of tolerances to achieve a proper model treatment. However, this does not mean that the task is trivial. As outlined in Section 2.3, the robust treatment of trimmed models is a really challenging issue in CAGD. Regarding IGA, an additional obstacle complicates the situation, that is, analysis software has to deal with extracted data. In other words, the input data provides only a reduced portion of the information that would be available in the initial CAD tool. Furthermore, this portion may be altered due to the translation process that might be required for the exchange. This aspect could be improved when the exchange procedure is tailored to a specific CAD system using its native data format. Yet, this would require vendor interaction and the restriction to a single software. Most importantly, this option is not very sustainable since a native format of a CAD system may become obsolete after a new software version is released.

### 3.2.2 Global approaches

Global approaches decompose trimmed model components into a set of regular surfaces or replace them by other spline representations such as subdivision surfaces or T-splines. Similar to the developments regarding local refinement, some strategies may originate from CAGD. For instance, isogeometric analysis with subdivision surfaces[61,62] and T-splines[36,63] can be included into the class of global techniques. Reconstruction concepts proposed in the context of analysis usually aim to replace trimmed surfaces by a set of regular ones. This is done by means of ruled surfaces[64], Coons patches[65], triangular Bézier patches[66], or a reconstruction based on isocurves[67].

Global approaches seek to resolve the core problem of trimmed solid models and hence, they are more related to CAGD than their local counterpart. Consequently, advances in this research area are more likely to have an impact in the design community. T-splines are a prime example in this regard. The introduction of T-splines in IGA has led to various enhancements such as analysis-suitable T-spline spaces that guarantee linear independent basis functions and it would be no exaggeration to say that IGA has been a driving force for the development T-splines in the past years. Approaches emerging from an analysis perspective can also be very useful for design applications. For instance, the reconstruction scheme introduced by Urick[67] could be utilized to create watertight Boolean operations. This possibility is currently under investigation and a preliminary example is illustrated in Figure 2. Note that Figure 2(c) shows a single surface with a matching parameterization across the computed intersection.

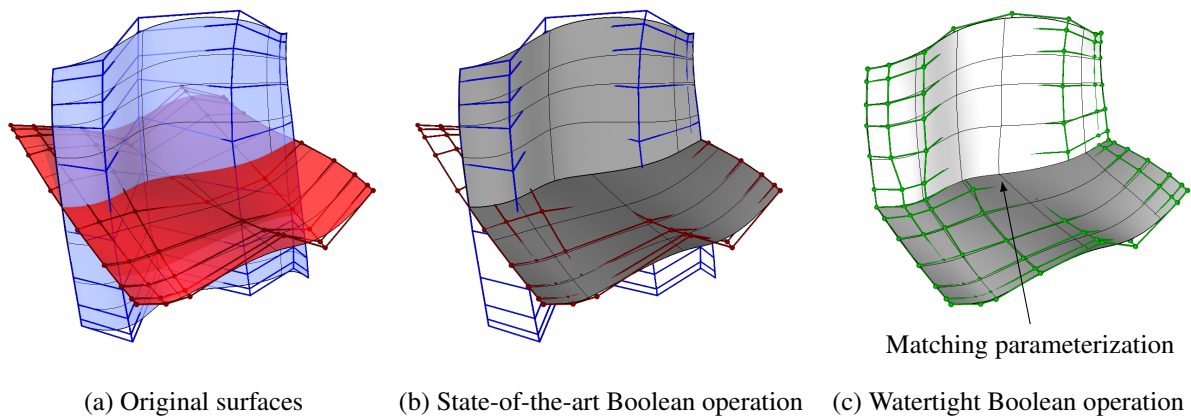

(a) Original surfaces    (b) State-of-the-art Boolean operation    (c) Watertight Boolean operation

Figure 2: Comparison of conventional and watertight Boolean operations: (a) initial surfaces and their control grids colored in blue and red, respectively, (b) the outcome based on a conventional Boolean operation, and (c) the result of the watertight counterpart. The control grids in (a) and (b) are identical, whereas in (c) the control points are updated to reflect the intersection which is defined as an isocurve of the watertight surface.

In contrast to local approaches, it is hard to identify general ingredients associated to global reconstruction procedures. Each global strategy requires a self-contained concept which becomes more and more sophisticated with its capabilities. This is indeed a potential drawback, especially when new features are added later on. Once a global scheme can be successfully applied, that is, it leads to a watertight model, two fundamental questions have to be addressed: (i) the representation of unstructured meshes and (ii) the treatment of extraordinary points (EPs). These topics are indeed of great interest for CAGD. Current model data is usually based on a structured mesh setting, where all control points of a surface are arranged in a regular grid. When local refinement of tensor product surfaces is considered as well, a structured mesh admits only interior points of valance 4 and T-junctions. But in case of smooth watertight models points with any valence (e.g., 3, 4, 5, ...) can occur and the arising non-regular points are referred to as EPs. These EPs also affect analysis properties and hence, their proper treatment is important for IGA[68]. It is worth noting that IGA researchers are also included in recent attempts seeking to include the capability of representing unstructured meshes in STEP.

Looking at the overall scope of global schemes, it is fair to say that they do address core issues of trimmed models. A compelling analysis-suitable approach could eventually resolve the robustness and interoperability of trimmed models not only for analysis, but all downstream applications. On the other hand, they are more complex and their success will also depend on their acceptance in CAGD.

## 4 CONCLUSION AND OUTLOOK

A brief overview of the development of surface-to-surface intersection operations and the formulation of trimmed solid models is provided to indicate the potential problems related to these popular computer aided geometric design (CAGD) representations. Strategies for isogemetric analysis (IGA) with trimmed geometries are listed and divided into two categories: (i) local approaches aim to enhance the analysis process and (ii) global approaches try to convert trimmed objects to regular models before the simulation. It is argued that these advances also bring new insights for CAGD, indicating the mutual benefits due to the interaction of the design and analysis communities. That IGA solutions lead to improvements for analysis as well as design has already been demonstrated by the evolution of local refinement concepts for multivariate splines and the recent developments regarding the treatment of trimmed models are indeed on a similar trajectory.